\documentclass[conference]{IEEEtran}
\IEEEoverridecommandlockouts
\usepackage[utf8]{inputenc}
\usepackage{cite}
\usepackage{cuted}
\usepackage{amsmath,amssymb,amsfonts}
\usepackage{bm}
\usepackage{cancel}
\usepackage[ruled,linesnumbered]{algorithm2e}
\usepackage{algorithmic}
\usepackage{graphicx}
\usepackage{textcomp}
\usepackage{xcolor}
\usepackage[export]{adjustbox}
\usepackage{xurl}
\usepackage{comment}
\usepackage{paralist} %for inparaenum
\usepackage{enumitem} %for short labels
\usepackage[T1]{fontenc}
\DeclareMathAlphabet{\mathpzc}{OT1}{pzc}{m}{it}
\usepackage{tabularx}
\usepackage{booktabs}
\usepackage{multirow}
\usepackage{mathtools}
\usepackage{makecell}
\usepackage{soul}
\usepackage{balance}
\usepackage{hhline}
\usepackage{booktabs}
\usepackage{enumitem}
\usepackage{subcaption}
\usepackage{caption}
\usepackage{float}

\usepackage{comment}

\usepackage[most]{tcolorbox}

\usepackage{nomencl}
\makenomenclature
\usepackage{etoolbox}
\renewcommand\nomgroup[1]{%
  \ifstrequal{#1}{P}{\vspace{10pt}\item[\textbf{Parameters}]}{%
  \ifstrequal{#1}{V}{\vspace{10pt}\item[\textbf{Variables}]}{}}{%
  \ifstrequal{#1}{S}{\vspace{10pt}\item[\textbf{Sets}]}{}}%
}

\usepackage{nccmath}
\usepackage{xpatch}
\xpatchcmd{\NCC@ignorepar}{%
\abovedisplayskip\abovedisplayshortskip}
{%
\abovedisplayskip0.2\abovedisplayshortskip%
\belowdisplayskip0.2\belowdisplayshortskip}
{}{}

\usepackage[belowskip=1pt,aboveskip=1pt]{caption}
\title{Stochastic Economic Dispatch with Battery Energy Storage considering Wind and Load Uncertainty}

\author{\IEEEauthorblockN{Shishir Lamichhane$^\dagger$, Anamika Dubey}
\IEEEauthorblockA{Washington State University, Pullman, Washington, USA}
Email: $^\dagger$shishir.lamichhane@wsu.edu
\thanks{This work is supported in part by the National Science Foundation (NSF)
under Career Award Number 1944142, and in part by the U.S. Department of
Energy’s Office of Energy Efficiency and Renewable Energy (EERE) under
the Solar Energy Technologies Office Award Number DE-EE00010424.}
\vspace{-0.4cm}}
\begin{document}
\bstctlcite{IEEEexample:BSTcontrol}
\maketitle
\thispagestyle{plain}
\pagestyle{plain}

\begin{abstract}

  With the integration of renewable energy resources in power systems, managing operational flexibility and reliability while minimizing operational costs has become increasingly challenging. Battery energy storage system (BESS) offers a promising solution to address these issues. This paper presents a stochastic dynamic economic dispatch with storage (SDED-S) framework to assess the impact of BESS in managing uncertainty. The temporal correlation between wind and load uncertainties is captured, with scenarios generated using a method inspired by stratified and importance sampling. The proposed approach is demonstrated on a modified IEEE 39-bus system, where selected conventional generators are converted to wind power plants.  Case studies show that strategic BESS deployment significantly improves system flexibility by reducing renewable curtailments and dispatch costs. Renewable energy curtailments decrease upon increasing BESS size and approach zero depending on wind penetration level. Higher wind penetrations result in greater curtailments without storage and yield larger cost savings when BESS is deployed, highlighting the growing need for flexibility as renewable energy resource's penetrations increase.
  % While cost savings increase with BESS capacity, diminishing returns are observed at larger sizes. These findings highlight that, beyond the value of BESS, optimal sizing of BESS is crucial to maximize economic gains and maintain reliability under high renewable integration.  
\end{abstract}


\section{Introduction}
% \begin{itemize}
% \ write uncertainty is increasing from renewables and laod both. give data of penetration. 
%\wite due to such high penetration and increasing uncertainty, economic dispatch should consider uncertainty. Furthermore, incorporation of BESS helps to improve the operational flexibility of the system.It helps reduce the curtailments and therby reducing the overall cost.  
%\ various solutions have been proposed to handle such situation
%\BESS is important 

% \end{itemize}

The penetration of renewable energy resources such as wind and solar has grown rapidly in recent years. According to the International Renewable Energy Agency, global installed renewable energy capacity surpassed 4000 GW by the end of 2024, driven by reduced installation costs and ambitious decarbonization goals worldwide~\cite{IRENA2025}. While these resources support clean-energy policy and reduce dependency on fossil fuels, they introduce operational challenges due to their inherent variability.  This uncertainty, coupled with rising demand from data centers, electric vehicles, further complicates grid operations. Additionally, the growing frequency of extreme weather events further heightens operational uncertainty during such times and poses large socio-economic effects~\cite{socio_economic_impact_abodh}. Ensuring secure and efficient grid operations under such stochastic conditions requires explicitly accounting for uncertainty in both planning and real-time operations.

In addition to increasing uncertainty, the integration of renewables by replacing conventional generators also reduces overall system flexibility due to fluctuations. This limits the full utilization of renewable energy and often leads to curtailments. 
% From the system operator's perspective, the variability of renewable generation reduces dispatchability, directly affecting the power system reliability. 
If these fluctuations can be mitigated, the ability to dispatch renewable power and integrate it effectively into the grid would improve significantly. To address these challenges, many utilities are turning to energy storage as a means to smooth renewable fluctuations and to cope with uncertainties, thereby increasing the reliability and reducing the operational costs. For instance, since 2021, the installed storage capacity of Electric Reliability Council of Texas (ERCOT) has increased by over 2400\%~\cite{powereng_texas_storage_2024}, and storage deployment on ERCOT has saved at least \$750 million since 2023~\cite{acp_ercot_storage_2024}. Therefore, the planning model should consider such storage while making decisions under uncertainty.

Day-ahead unit commitment models typically operate over a 24-hour horizon with hourly resolution, using forecasted values to determine generation schedules and reserves. However, forecast accuracy declines with longer horizons; for example, wind forecast errors can increase by up to 35\%~\cite{PH_GU_wind_load_only}. These errors are expected to grow with higher renewable penetration and rising demand from electric vehicles and data centers. Therefore, incorporating uncertainty in near real-time dispatch, along with energy storage modeling, is essential to increase the reliability and reduce the operating costs.

A chance-constrained economic dispatch model with storage is proposed in~\cite{energy_storage_chance_constraint}. However, such chance constraint models require prior knowledge of probability distributions and can often lead to non-convex formulations even if the original problem is convex~\cite{LineRoaldReviewStochastic}. A robust optimization approach has been implemented in~\cite{robust_optimization_storage}, which ensures feasibility in all scenarios, but it leads to overly conservative and costly decisions. In contrast, stochastic optimization addresses uncertainty by considering a set of scenarios and determining recourse actions once a scenario is realized. This approach does not require explicit modeling of probability distributions, and forecasted data can be directly used as scenarios~\cite{LineRoaldReviewStochastic}. Therefore, we adopt a stochastic optimization framework in this paper.

A stochastic dynamic economic dispatch model (SDED) is presented in~\cite{scenario_set_ED_gen_parameters}, incorporating storage and wind power, where the problem is solved using a particle swarm optimization algorithm. The study analyzes the impact of storage size under varying wind penetration levels, with scenarios obtained using a k-means clustering algorithm. In~\cite{Linli_SDED_storage_PSO_bacterial}, an economic dispatch model considering wind uncertainty and BESS is solved using a metaheuristic approach. A very short-term SDED model is developed in~\cite{stochastic_SDED_FR}, incorporating wind power and BESS to schedule flexible ramping reserves, with scenarios generated via a point estimation method. However, these studies either do not explore the combined impact of varying renewable penetration levels and BESS sizes, neglect the temporal correlation between uncertain parameters, overlook curtailment analysis, or do not provide an explicit two-stage modeling framework for SDED with storage. 

In this paper, we address these gaps by developing a two-stage stochastic dynamic economic dispatch model with storage (SDED-S) framework. We generate scenarios using a method inspired by stratified and importance sampling, utilizing probabilistic distribution data while preserving the temporal correlation between wind and load. We then conduct a comprehensive analysis by varying both wind penetration levels and BESS sizes, solving the extensive form to evaluate impacts on curtailments, dispatch decisions, and dispatch costs. The rest of the paper is organized as follows: Section~\ref{sec:Modeling} describes the BESS model, the two-stage SDED-S formulation, and uncertainty modeling. Section~\ref{sec:results and discussion} presents results and discussion, followed by the conclusion in section~\ref{sec:Conclusion}.

\section{Modeling}\label{sec:Modeling}

\subsection{Modeling of Battery Energy Storage System (BESS)} \label{subsec: battery modeling}

In this section, we present the BESS model, which is integrated into the economic dispatch framework in the coming sections. Let $P^{ch}_t$ and $P^{dis}_t$ denote the charging and discharging power of the battery at time $t$, respectively. Similarly, let $\gamma^{ch}_t$ and $\gamma^{dis}_t$ be binary variables indicating whether the battery is in charging or discharging mode at time $t$. The control signals for the battery must satisfy the following logic conditions in addition to operational limit constraints. The final logic constraint indicates that the battery cannot charge and discharge simultaneously.

\begin{center}
\begin{itemize}
    \item If $\gamma^{ch}_t = 1$, then $P^{ch}_t > 0$
    \item If $\gamma^{dis}_t = 1$, then $P^{dis}_t > 0$
    \item $\gamma^{ch}_t \gamma^{dis}_t = 0$
\end{itemize}
\end{center}

These logical, along with limit-based conditions, are modeled using the big-M method in (\ref{eq:battery charging limit}) - (\ref{eq: SOC limit}). Constraint~(\ref{eq:battery charging limit})  ensures that $P^{ch}_t$ does not exceed the maximum charging rate  $\overline{P^{ch}}$ when $\gamma^{ch}_t$ is 1, and is zero otherwise. A similar formulation applies to discharging in (\ref{eq:battery discharging limit}), with $\overline{P^{dis}}$ representing the maximum discharging rate. Charging and discharging of the battery, both at the same time, are avoided by (\ref{eq:battery complementarity}). The evolution of the state of charge (SOC) of the battery is modeled by~(\ref{eq:battery SOC}), where $E^{cap}$ is the energy capacity of the battery, $\Delta t$ is the dispatch interval duration, and $\eta^{ch}$, $\eta^{dis}$ are the charging and discharging efficiencies, respectively. Finally, (\ref{eq: SOC limit}) bounds the SOC within its minimum and maximum permissible limits. 

\begin{equation}
    0 \leq P^{ch}_t \leq \gamma^{ch}_t \overline{P^{ch}} \label{eq:battery charging limit}
\end{equation}

\begin{equation}
    0  \leq P^{dis}_t \leq \gamma^{dis}_t\overline{P^{dis}} \label{eq:battery discharging limit}
\end{equation}

\begin{equation}
    \gamma^{ch}_t + \gamma^{dis}_t \leq 1 \label{eq:battery complementarity}
\end{equation}

\begin{equation}
    SOC_{t} = SOC_{t-1} + \left(\eta^{ch}P^{ch}_t - \frac{P^{dis}_t}{\eta^{dis}}\right) \frac{\Delta t}{E^{cap}} \label{eq:battery SOC}
\end{equation}

\begin{equation}
    \underline{SOC} \leq SOC_t \leq \overline{SOC} \quad \label{eq: SOC limit}
\end{equation}

\subsection{Stochastic Dynamic Economic Dispatch with Storage (SDED-S)} 

We model SDED-S as a two-stage stochastic optimization problem, building upon the SDED model introduced in~\cite{lamichhane2025comparison}. Generators and battery dispatch decisions are made in the first stage based on forecasted information for load and wind power. The second stage then determines the recourse actions, such as generator regulations, battery redispatch, and curtailments, once the uncertainty is realized. The power system is represented as a graph $\mathbb{G}(\mathcal{N}, \mathcal{L})$, where $\mathcal{N}$ is the set of buses and $\mathcal{L}$ is the set of lines. We adopt the DC power flow approximation, a widely used and computationally efficient method for economic dispatch. 

\subsubsection{First Stage}

Let $G$, $G^w$, and $B$ denote the sets of conventional generators, wind plants, and BESS units, respectively. The set of dispatch intervals is denoted by $\mathcal{T}$. The first-stage objective function~(\ref{eq: first stage SDED obj}) minimizes the total generation cost, battery dispatch cost, curtailment costs, and the expected second-stage cost. The function  $e^g(.)$ represents the generation cost of generator  $g \in G$, and $x_{g,t}$ is the corresponding output power at time $t$. Battery charging and discharging power for storage unit $b \in B$ at time $t$ are denoted by  $x^{ch}_{b,t}$  and $x^{dis}_{b,t}$, respectively. The parameters $c^{ch}$ and $c^{dis}$ denote the cost rates for battery charging and discharging, respectively. Curtailments are allowed to maintain the feasibility under uncertainty. Specifically, $\hat{x}_{g,t}$, $\hat{d}_{i,t}$, and $\hat{u}_{w,t}$ represent curtailments of generator $g$, load at bus $i$, and wind power plant at plant $w$, respectively at time $t$. These are penalized in the objective function using large penalty coefficients  $c^g$, $c^l$, and $c^w$, which correspond to generator, load, and wind curtailments, respectively. The final term in the objective represents the expected second-stage cost, which is detailed in Section \ref{secsec:second stage}.

\begin{equation}
\begin{gathered}
\min \sum_{g\in G}\sum_{t \in \mathcal{T}}[e^g(x_{g,t}) +c^g\hat{x}_{g,t}]  + \sum_{i \in \mathcal{N}} \sum_{t \in \mathcal{T}}c^l \hat{d}_{i,t}\\ 
+\sum_{b\in B}\sum_{t \in \mathcal{T}}[c^{ch}x^{ch}_{b,t} + c^{dis}x^{dis}_{b,t}] \\ 
+ \sum_{w \in G^w} \sum_{t \in \mathcal{T}} c^w \hat{u}_{w,t} +  \mathbb{E}[Q(\mathbf{x},\zeta)]
    \end{gathered}
    \label{eq: first stage SDED obj}
\end{equation}

The first-stage constraints are defined from (\ref{eq: first stage SDED power balance}) - (\ref{eq: first stage SOC limit}). Let  $G_i$, $G_i^w$ and $B_i$ represent the sets of conventional generators, wind plants, and BESS units connected to bus $i$, respectively. Let $p_{ij,t}$  denote the power flow from bus $i$ to bus $j$ through line $ij$ at time $t$. The power balance constraint (\ref{eq: first stage SDED power balance}) ensures the power balance, where forecasted wind power output $\overline{u}_{w,t}$ for plant $w$ and forecasted load demand $\overline{d}_{i,t}$ at bus $i$ for time $t$ are considered.

\begin{equation}
\begin{gathered}
    \sum_{g \in G_i} (x_{g,t} - \hat{x}_{g,t}) + \sum_{w \in G_i^{w}} (\overline{u}_{w,t} - \hat{u}_{w,t})\\ + \sum_{b \in B_i} (x^{dis}_{b,t}  - x^{ch}_{b,t}) + \sum_{j:(j,i)\in \mathcal{L}} p_{ji,t} =\\ \sum_{j:(i,j)\in \mathcal{L}} p_{ij,t}  + \overline{d}_{i,t} - \hat{d}_{i,t}  \ \  \forall i \in \mathcal{N}, \  \forall t \in \mathcal{T} \label{eq: first stage SDED power balance}
\end{gathered}
\end{equation}

The power flow through line $ij$ is modeled by (\ref{eq: first stage SDED angle with line power}), where $B_{ij}$ denotes the susceptance of line $ij$, $\delta_{i,t}$ is the voltage angle at bus $i$ at time $t$, and $\delta_{ij}^{sh}$ represents the phase shift angle associated with the line. The power flow limit is enforced by (\ref{eq: first stage SDED transmission line limit}), where $\overline{P_{ij}}$ is its maximum thermal limit. Additionally, the voltage angle difference across the line $ij$ is bounded between $\underline{\delta_{ij}}$ and $\overline{\delta_{ij}}$, as specified by (\ref{eq: first stage SDED angle difference limit}).

\begin{equation}
    p_{ij,t} = B_{ij} (\delta_{i,t} - \delta_{j,t} - \delta_{ij}^{sh}) \ \  \forall ij \in \mathcal{L}, \  \forall t \in \mathcal{T}  \label{eq: first stage SDED angle with line power}
\end{equation}

\begin{equation}
    -\overline{P_{ij}} \leq p_{ij,t} \leq \overline{P_{ij}} \ \  \forall ij \in \mathcal{L}, \  \forall t \in \mathcal{T}  \label{eq: first stage SDED transmission line limit}
\end{equation}

\begin{equation}
    \underline{\delta_{ij}} \leq \delta_{i,t} - \delta_{j,t} \leq \overline{\delta_{ij}} \ \  \forall ij \in \mathcal{L}, \  \forall t \in \mathcal{T} \label{eq: first stage SDED angle difference limit}
\end{equation}

Let $R_g^{down}$ and $R_g^{up}$ denote the ramp-down and ramp-up limits of generator $g$, respectively, which are determined based on ramp rate and the dispatch interval duration. The output from the generator is constrained between consecutive time intervals by the ramping limits as specified in ($\ref{eq: first stage SSED ramp limit}$). Additionally, the dispatch of generator $g$ is bounded between its minimum generation limit   $\underline{x}_g$ and maximum generation capacity $\overline{x}_g$.

\begin{equation}
    -R_g^{down} \leq x_{g,t} - x_{g,t-1} \leq R_g^{up} \ \  \forall g \in G, \  \forall t \in \mathcal{T} \label{eq: first stage SSED ramp limit}
\end{equation}

\begin{equation}
    \underline{x}_g \leq x_{g,t} \leq \overline{x}_g \ \  \forall g \in G, \  \forall t \in \mathcal{T} \label{eq:first stage SDED limit generator}
\end{equation}

Let  $\gamma^{ch}_{b,t}$ and $\gamma^{dis}_{b,t}$ be binary variables indicating whether BESS unit $b$ is charging or discharging at time $t$, respectively. These variables determine the battery's operational mode based on the logic outlined in Section~\ref{subsec: battery modeling}. The charging and discharging power of the battery are constrained by (\ref{eq: first stage battery charging limit}) and (\ref{eq: first stage battery discharging limit}), respectively, where $\overline{x^{ch}_b}$ and $\overline{x^{dis}_b}$ represent the maximum charging and discharging capacities of the storage unit. Constraint (\ref{eq: first stage battery complementarity}) ensures that the battery cannot charge and discharge simultaneously. Note that the battery can remain idle during a time period, in which case both binary variables take zero value. The evolution of the state of charge of the battery is governed by (\ref{eq: first stage battery SOC}). Finally, (\ref{eq: first stage SOC limit}) constraints the SOC of the battery within minimum and maximum limits at all time intervals $t$.

\begin{equation}
    0 \leq x^{ch}_{b,t} \leq \gamma^{ch}_{b,t} \overline{x^{ch}_b}\ \   \forall b \in B, \  \forall t \in \mathcal{T}  \label{eq: first stage battery charging limit}
\end{equation}

\begin{equation}
   0 \leq x^{dis}_{b,t} \leq \gamma^{dis}_{b,t} \overline{x^{dis}_b}\ \   \forall b \in B, \  \forall t \in \mathcal{T}  \label{eq: first stage battery discharging limit}
\end{equation}

\begin{equation}
    \gamma^{ch}_{b,t} + \gamma^{dis}_{b,t} \leq 1  \ \   \forall b \in B, \  \forall t \in \mathcal{T}  \label{eq: first stage battery complementarity}
\end{equation}

\begin{equation}
    SOC_{b,t}\!=\! SOC_{b,t-1}\! +\! \!\left(\eta^{ch}x^{ch}_{b,t}\! - \!\frac{x^{dis}_{b,t}}{\eta^{dis}}\right)\! \!\frac{\Delta t}{E^{cap}_b} \   \forall b\! \in\! B,  \forall t\! \in\! \mathcal{T} \label{eq: first stage battery SOC}
\end{equation}

\begin{equation}
    \underline{SOC} \leq SOC_{b,t} \leq \overline{SOC}   \ \   \forall b \in B, \  \forall t \in \mathcal{T}  \label{eq: first stage SOC limit}
\end{equation}

\subsubsection{Second Stage} \label{secsec:second stage}

After determining the generator dispatch decisions in the first stage, recourse actions must be taken in the second stage once the uncertainty is realized. These recourse actions aim to ensure system feasibility while minimizing the overall operating cost. All parameters and variables associated with a scenario $\zeta$ will henceforth be indexed by $\zeta$. Let $\mathbf{x}$ represent the set of first-stage dispatch decision variables, which serve as inputs to the second-stage problem. Let $G^{R}$ denote the set of generators capable of providing regulation services. The variables $x_{g,t,\zeta}^+$ and  $x_{g,t,\zeta}^-$ represent the up-regulation and down-regulation power from generator $g \in G^R$ at time $t$ under scenario $\zeta$, respectively. The second-stage objective function is given by  (\ref{eq:SDED second stage objective}), which minimizes the total cost associated with regulation, battery operations (charging and discharging), and curtailments. The parameters $r_g^+$ and $r_g^-$ denote the cost rate of up-regulation and down-regulation for generator $g$, respectively. Other parameters and variables retain the same definition as in the first stage, except they are now indexed by $\zeta$. 

\begin{equation}
\begin{gathered}
Q(\mathbf{x},\zeta) = \min \sum_{g \in G^{R}} \sum_{t \in \mathcal{T}} [r_g^+x_{g,t,\zeta}^+ + r_g^-x_{g,t,\zeta}^-]\\
+ \sum_{w \in G^W} \sum_{t \in \mathcal{T}} c^{w}\hat{u}_{w,t,\zeta}  +\sum_{b\in B}\sum_{t \in \mathcal{T}}[c^{ch}x^{ch}_{b,t, \zeta} + c^{dis}x^{dis}_{b,t,\zeta}] \\ 
+  \sum_{i \in \mathcal{N}} \sum_{t \in \mathcal{T}} c^{l}\hat{d}_{i,t,\zeta} + \sum_{g \in G} \sum_{t \in \mathcal{T}} c^{g}\hat{x}_{g,t,\zeta} 
\end{gathered}
\label{eq:SDED second stage objective}
\end{equation}

The second-stage constraints are defined by (\ref{eq:SSED second stage power balance})- (\ref{eq: second stage SOC limit}). Let  $u_{w,t,\zeta}$ denote the actual wind power output from wind plant $w$, $d_{i,t,\zeta}$ represents the realized load demand at bus $i$, both at time $t$ under scenario $\zeta$. Power balance at each bus and time interval is enforced by~(\ref{eq:SSED second stage power balance}), where $G_i^{R}$ is the set of regulation generators connected to bus $i$. It is important to note that if a regulation generator does not participate in the first-stage dispatch, its scheduled output  $x_{g,t}$ is set to zero in the second stage.

\begin{equation}
\begin{gathered}
    \sum_{g \in G_i \setminus G_i^{R}} (x_{g,t} - \hat{x}_{g,t,\zeta}) + \sum_{g \in G_i^{R}} (x_{g,t} + x_{g,t,\zeta}^+ - x_{g,t,\zeta}^-)\\  + \sum_{w \in G_i^{w}}(u_{w,t,\zeta} -\hat{u}_{w,t,\zeta}) +\sum_{b\in B}[c^{ch}x^{ch}_{b,t,\zeta} + c^{dis}x^{dis}_{b,t,\zeta}]\\
    +  \sum_{j:(j,i)\in \mathcal{L}} p_{ji,t,\zeta}
    - \sum_{j:(i,j)\in \mathcal{L}} p_{ij,t,\zeta}\\ -  (d_{i,t,\zeta} - \hat{d}_{i,t,\zeta}) = 0 \   \forall i \in \mathcal{B},  \forall t \in \mathcal{T} 
    \end{gathered}
     \label{eq:SSED second stage power balance}
\end{equation}

Constraints (\ref{eq:SSED second stage regulating up variable}) and (\ref{eq:SSED second stage regulating down variable}) ensure that the total generation from a regulation generator $g$ after applying up-regulation $x_{g,t,\zeta}^+$ and down-regulation  $x_{g,t,\zeta}^-$, remains within its allowable generation limits. The ramping constraint following regulation actions between two consecutive time intervals is enforced by (\ref{eq:SSED second stage ramp limit}). Equations  (\ref{eq:SSED angle with line power second stage}) and (\ref{eq:SSED second stage line limit}) model and limit the power flow for line $ij$ at time $t$ during the realization of scenario $\zeta$, respectively. Additionally, (\ref{eq:SSED second stage angle difference limit}) constraints the voltage angle difference across the line within limits. The remaining constraints (\ref{eq: second stage battery charging limit}) - (\ref{eq: second stage SOC limit}) are related to battery storage units. These constraints follow the same structure and logic as in the first stage, with the key difference that they are now scenario-dependent indexed by $\zeta$.

\begin{equation}
    x_{g,t} + x_{g,t,\zeta}^+ \leq \overline{x}_g \ \  \forall g \in G^{R}, \ \forall t \in \mathcal{T} \label{eq:SSED second stage regulating up variable}
\end{equation}

\begin{equation}
    x_{g,t}  - x_{g,t,\zeta}^-  \geq \underline{x}_g \ \  \forall g \in G^{R},  \  \forall t \in \mathcal{T} \label{eq:SSED second stage regulating down variable}
\end{equation}

\begin{equation}
\begin{gathered}
    -R_g^{down} \leq x_{g,t} + x_{g,t,\zeta}^+ - x_{g,t,\zeta}^- - x_{g,t-1}\\
    - x_{g,t-1,\zeta}^+ + x_{g,t-1,\zeta}^- \leq R_g^{up} \ \  \forall g \in G^{R}, \  \forall t \in \mathcal{T}
    \end{gathered} \label{eq:SSED second stage ramp limit}
\end{equation}

\begin{equation}
    p_{ij,t,\zeta} = B_{ij} (\delta_{i,t,\zeta} - \delta_{j,t,\zeta} - \delta_{ij}^{sh}) \ \  \forall ij \in \mathcal{L}, \  \forall t \in \mathcal{T} \label{eq:SSED angle with line power second stage}
\end{equation}

\begin{equation}
\begin{gathered}
    -\overline{P_{ij}} \leq p_{ij,t,\zeta} \leq \overline{P_{ij}} \ \  \forall ij \in \mathcal{L}, \  \forall t \in \mathcal{T}
    \end{gathered} \label{eq:SSED second stage line limit}
\end{equation}

\begin{equation}
    \underline{\delta_{ij}} \leq \delta_{i,t,\zeta} - \delta_{j,t,\zeta} \leq \overline{\delta_{ij}} \ \  \forall ij \in \mathcal{L}, \  \forall t \in \mathcal{T} \label{eq:SSED second stage angle difference limit}
\end{equation}

\begin{equation}
    0 \leq x^{ch}_{b,t, \zeta} \leq \gamma^{ch}_{b,t, \zeta} \overline{x^{ch}_b}\ \   \forall b \in B, \  \forall t \in \mathcal{T}  \label{eq: second stage battery charging limit}
\end{equation}

\begin{equation}
   0 \leq x^{dis}_{b,t, \zeta} \leq \gamma^{dis}_{b,t, \zeta} \overline{x^{dis}_b}\ \   \forall b \in B, \  \forall t \in \mathcal{T}  \label{eq: second stage battery discharging limit}
\end{equation}

\begin{equation}
    \gamma^{ch}_{b,t, \zeta} + \gamma^{dis}_{b,t, \zeta} \leq 1  \ \   \forall b \in B, \  \forall t \in \mathcal{T}  \label{eq: second stage battery complementarity}
\end{equation}

\begin{equation}
\begin{gathered}
    SOC_{b,t,\zeta} = SOC_{b,t-1, \zeta} + \left(\eta^{ch}x^{ch}_{b,t, \zeta} - \frac{x^{dis}_{b,t, \zeta}}{\eta^{dis}}\right) \frac{\Delta t}{E^{cap}}  \\  \quad \quad \forall b \in B, \  \forall t \in \mathcal{T}  
    \end{gathered}
    \label{eq: second stage battery SOC}
\end{equation}

\begin{equation}
    \underline{SOC} \leq SOC_{b,t, \zeta} \leq \overline{SOC}  \ \   \forall b \in B, \  \forall t \in \mathcal{T}  \label{eq: second stage SOC limit}
\end{equation}

\subsection{Load and Wind Uncertainty Modeling} \label{scenario generation}

We obtain the sub-hourly wind and load forecast data from ARPA-E PERFORM dataset developed by the National Renewable Energy Laboratory (NREL)~\cite{OEDI_Dataset_5772}. Load forecasts are generated by using a deep learning ensemble that combines recurrent neural networks, convolutional neural networks, and extreme gradient boosting, while wind forecasts are produced using a machine learning-based multi-model framework with historical synthetic data. These deterministic forecasts are converted into probabilistic forecasts using an adaptive Gaussian model~\cite{OEDI_Dataset_5772}. The dataset includes $99^{th}$ percentile values, which are first normalized, and the cumulative distribution function (CDFs) are constructed. Then, discrete probability density functions (PDFs) are derived for each sub-hourly step. After this, based on the required number of scenarios, the number of strata and the data points within each stratum are identified, and different strata are formed for each PDF, where each stratum represents a specific equal-probability segment of the cumulative distribution. Then, data within each stratum is averaged to obtain a representative scenario, and the corresponding probability is updated accordingly. This similar approach, motivated by stratified and importance sampling, has also been employed in~\cite{poudyal2023resiliencedrivenplanningelectricpower}. The resulting system-level wind and load multipliers are then spatially distributed to bus-level scenarios by adding Gaussian noise to introduce spatial variability, as utilized in~\cite{lamichhane2025SPAR}.

\section{Results and Discussions} \label{sec:results and discussion}

\begin{figure}[t]
    \centering
    \includegraphics[width=1\linewidth]{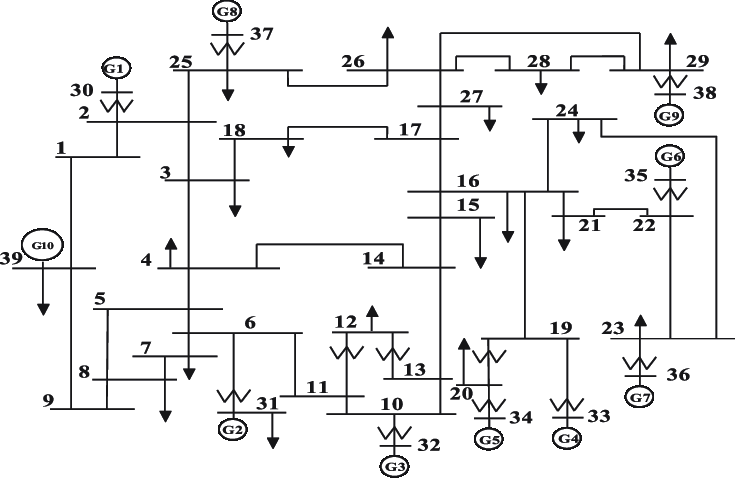}
    \caption{IEEE 39-bus system.}
    \label{fig:39 bus system}
\end{figure}

The proposed framework is demonstrated using the IEEE-39 bus system as shown in  Figure~\ref{fig:39 bus system}, where selected conventional generators are replaced with wind power plants of equivalent capacity. The system comprises 39 buses, 46 branches, and 10 generators, with a total system load of 6254.2 MW and an aggregate generation capacity of 7367 MW. Generator details are summarized in Table~\ref{tab:generator details}, including their respective cost parameters, which are obtained from \cite{scenario_set_ED_gen_parameters}. In the table,  $\overline{x}_g$ and $\underline{x}_g$ represent the maximum and minimum generation limits, while  $a$, $b$, and $c$ denote the coefficients of the quadratic generation cost function. The final column, ``\textit{Ramp}" specifies the generation ramping rate in MW per minute.

To perform a sensitivity analysis, we consider different levels of wind penetration by modifying the combination of the generators. Wind penetration is defined as the ratio of total installed wind power capacity to the total generation capacity of the system. Specifically, converting generator $G3$ to a wind power plant results in a 10\% penetration level. Replacing both   $G3$ and $G6$ yields 20\% penetration, while replacing $G3$, $G4$, and $G6$ results in approximately 30\% penetration. A 40\% penetration level is achieved by converting $G3$, $G4$, $G6$, and $G9$ to wind power plants. Note that the wind plant locations are not optimized but are selected by replacing existing generators to meet desired wind penetration levels while maintaining the system structure. Although cost coefficients and ramping capabilities are listed for all generators in Table~\ref{tab:generator details}, these values are not applicable if a generator is replaced by a wind power plant. In such cases, the corresponding cost coefficients and ramping capacity are assumed to be zero.  The default configuration used for simulation assumes a 20\% wind penetration level. In addition, one BESS unit is installed at each of buses 21 and 28. The BESS unit size is varied from 20 MW (with 4-hour duration capacity) to 120 MW (4-hour duration capacity) to evaluate different storage scenarios and their impact on the system performance. Since BESS units are not part of the original system,  their locations are selected near wind installations in this work, and optimizing their placements is left as a potential direction of future work.

The regulation cost for each generator is assumed to be 1.5 times its normal linear generation cost. All conventional generators are assumed to be capable of providing regulation services. The cost parameters for various system operations are as follows: wind power curtailment cost ($c^w$) is set at \$100/MWh, load curtailment cost ($c^l$) at \$3000/MWh, and conventional generation curtailment cost ($c^g$) at \$400/MWh. The costs for battery charging($c^{ch}$) and discharging($c^{dis}$) are both assumed to be \$10/MWh. The SOC for the battery is constrained between a minimum of 10\% and a maximum of 90\%. The initial SOC level is considered to be 50\%. It is important to note that all parameter values used in this study can be adjusted within the proposed framework to reflect different utility practices or policy requirements.

We obtain load and wind power multipliers from the western region of the ERCOT using PERFORM datasets produced by NREL~\cite{OEDI_Dataset_5772}. The dispatch horizon spans 2 hours with a 15-minute interval, resulting in 8 time intervals starting from 3 pm, January 04, 2018 UTC. After obtaining the system-level load and wind multipliers, 10\% Gaussian noise is added to obtain bus-level load and wind multipliers. We generate 50 scenarios using the method outlined in Section \ref{scenario generation}. The optimization model was formulated using Pyomo\cite{pyomo} and was solved using Gurobi~\cite{gurobi} on a workstation equipped with an Intel Core i9-10900X CPU (3.7GHz) and 32 GB of RAM.

% \begin{table}[t]
% \small
% \centering
% \setlength{\tabcolsep}{5.5pt}
% \caption{Generators Details}
% \label{tab:generator details}
% \begin{tabularx}{\linewidth}{ccccccc}
% \hline
% \textit{Generators} & \textit{\begin{tabular}[c]{@{}c@{}}\overline{x}_g\\ MW\end{tabular}} & \textit{\begin{tabular}[c]{@{}c@{}}\underline{x}_g\\ MW\end{tabular}} & \textit{\begin{tabular}[c]{@{}c@{}}a\\ \$/MW^2\end{tabular}} & \textit{\begin{tabular}[c]{@{}c@{}}b\\ \/MW\end{tabular}} & \textit{\begin{tabular}[c]{@{}c@{}}c\\ \$\end{tabular}} & \textit{\begin{tabular}[c]{@{}c@{}}Ramp\\  MW/Min\end{tabular}} \\ \hline
% G1 & 1040 & 0 & 0.00048 & 16.19 & 1000 & 6.2 \\ \hline
% G2 & 646 & 0 & 0.00031 & 17.26 & 970 & 3.8 \\ \hline
% G3 & 725 & 0 & 0.00211 & 16.5 & 680 & 4.3 \\ \hline
% G4 & 652 & 0 & 0.002 & 16.6 & 700 & 3.9 \\ \hline
% G5 & 508 & 0 & 0.00398 & 19.7 & 450 & 3.1 \\ \hline
% G6 & 687 & 0 & 0.00712 & 22.26 & 370 & 4.11 \\ \hline
% G7 & 580 & 0 & 0.00079 & 27.74 & 480 & 3.5 \\ \hline
% G8 & 564 & 0 & 0.00413 & 25.92 & 660 & 3.4 \\ \hline
% G9 & 865 & 0 & 0.00222 & 27.27 & 665 & 5.2 \\ \hline
% G10 & 1100 & 0 & 0.00173 & 27.79 & 670 & 6.6 \\ \hline
% \end{tabularx}
% \end{table}

\begin{table}[t]
\small
\centering
\setlength{\tabcolsep}{5.5pt}
\caption{Generators Details}
\label{tab:generator details}
\begin{tabularx}{\linewidth}{ccccccc}
\hline
\textit{Generators} &
\textit{\begin{tabular}[c]{@{}c@{}}$\overline{x}_g$\\ MW\end{tabular}} &
\textit{\begin{tabular}[c]{@{}c@{}}$\underline{x}_g$\\ MW\end{tabular}} &
\textit{\begin{tabular}[c]{@{}c@{}}$a$\\ \$/\(\mathrm{MW}^2\)\end{tabular}} &
\textit{\begin{tabular}[c]{@{}c@{}}$b$\\ \$/\(\mathrm{MW}\)\end{tabular}} &
\textit{\begin{tabular}[c]{@{}c@{}}$c$\\ \$\end{tabular}} &
\textit{\begin{tabular}[c]{@{}c@{}}Ramp\\ MW/min\end{tabular}} \\ \hline
G1  & 1040 & 0 & 0.00048 & 16.19 & 1000 & 6.2 \\ \hline
G2  &  646 & 0 & 0.00031 & 17.26 &  970 & 3.8 \\ \hline
G3  &  725 & 0 & 0.00211 & 16.50 &  680 & 4.3 \\ \hline
G4  &  652 & 0 & 0.00200 & 16.60 &  700 & 3.9 \\ \hline
G5  &  508 & 0 & 0.00398 & 19.70 &  450 & 3.1 \\ \hline
G6  &  687 & 0 & 0.00712 & 22.26 &  370 & 4.11\\ \hline
G7  &  580 & 0 & 0.00079 & 27.74 &  480 & 3.5 \\ \hline
G8  &  564 & 0 & 0.00413 & 25.92 &  660 & 3.4 \\ \hline
G9  &  865 & 0 & 0.00222 & 27.27 &  665 & 5.2 \\ \hline
G10 & 1100 & 0 & 0.00173 & 27.79 &  670 & 6.6 \\ \hline
\end{tabularx}
\end{table}

% \begin{itemize}
% \item \textcolor{red}{Give table which gives data for generators and also mention that wind penetration of 20\% means these these generators are assumed to be wind. Also give cost parameters utilized for each generators for regulation cost. }
%     \item \textcolor{red}{For wind penetration fo 20\%, give table which shows wind curtailment, cost for each battery size. Column will be battery size, objective function, improvement percentage}
%     \item \textcolor{red}{Give a plot which shows the dispatch pattern of two batteries in each time step, maybe merge into one plot for dispatch value if looks good., also give dispatch of wind values at 20\% generation for  bus 2 and 5., then explain the battery curve after showing wind power and load values, no need to show curve for generation dispatch. But you may show barchart for remaining 8 generators in long figure if space permits for each time period; otherwise no need to show dispatch of each generator as it is not required to explain other things. }
%     \item \textcolor{red}{Then for each penetration level of wind, plot \% improvement in dispatch cost with the increase in battery size. Its a good plot which gives good information. Maybe to explain why there is sharp increase in cost for 40\% penetration, give line flows in expected case with its capacity in diagram and compare it for with 20\%.}
% \end{itemize}

Table~\ref{tab:wind_penetration_20__battery_size} presents the dispatch cost for a system with 20\% wind penetration, where the size of each BESS is varied from 0 MW to 120 MW. The column labeled ``Cost Savings" indicates the savings amount achieved by installing the BESS of the specified size (shown in the first column), relative to the case with no BESS. The percentage in parentheses represents the relative percentage savings. It can be observed that as the BESS size increases from 0 MW to 60 MW, the dispatch costs decrease consistently, with savings improving from 3.14\% to 7.78\%. However, beyond 60 MW, the additional savings diminish, only increasing by about 2.3\% from 60 MW to 120 MW. This indicates a saturation effect, where increasing the storage capacity beyond a certain point yields limited economic benefits. These results highlight the importance of optimal BESS sizing to maximize cost-effectiveness under a given wind penetration level.

\begin{table}[t]
\small
\centering
\setlength{\tabcolsep}{16pt}
\caption{Results For 20\% Wind Penetration With Varying BESS Size}
\label{tab:wind_penetration_20__battery_size}
\begin{tabularx}{\linewidth}{ccc}
\hline
\textbf{\begin{tabular}[c]{@{}c@{}}\textit{BESS Size (MW)}\end{tabular}} &
\textbf{\begin{tabular}[c]{@{}c@{}}\textit{Cost(\$)}\end{tabular}} &
\textbf{\begin{tabular}[c]{@{}c@{}}\textit{Cost Savings (\%)}\end{tabular}} \\ \hline
0     & 191836 & -- \\ \hline
20   & 185798 & 6038 (3.14\%) \\ \hline
40   & 180748 & 11088 (5.77\%) \\ \hline
60      & 176906 & 14930 (7.78\%) \\ \hline
80       & 175099 & 16737 (8.72\%) \\ \hline
100     & 173616 & 18220 (9.49\%) \\ \hline
120     & 172448 & 19388 (10.1\%) \\ \hline
\end{tabularx}
\end{table}

Table~\ref{tab:dispatch values} summarizes the expected demand, dispatch, and curtailments across eight time intervals (T1-T8) under a case with 20\% wind penetration and  20 MW of BESS unit, resulting in a total of 40 MW BESS capacity in the system. The ``Demand" column indicates the system load, ``CGD" represents the dispatch from conventional generators, ``WG" is the available wind generation, ``BD" denotes the total dispatch from BESS, and ``WC" shows the wind curtailments. Load and conventional generation curtailments are not reported, as they were zero for all intervals. To minimize the total dispatch cost, the BESS is fully discharged to its full 40 MW capacity from T1 to T4. In T5, even though the net load increases slightly,  the battery dispatch is limited to 30.4 MW instead of discharging fully, to conserve energy for T6. This is because in T6, demand rises sharply to 4399.4 MW and the wind generator drops to 371.8 MW. Given that the ramping capability of conventional generators alone could not meet this increased net load, the battery was required to dispatch at full 40 MW capacity in T6 to maintain system balance and minimize costs. Between T6 and T7, demand drops significantly by 796 MW, but due to limited ramp-down capability, wind curtailment occurs. The BESS charges at its full 40 MW capacity (negative sign indicates charging) in T7, reducing wind curtailment to 146.8 MW, which otherwise was found to be 226 MW without storage. These results highlight how proactive and strategic BESS operation helps manage variability in system while minimizing curtailments.

\begin{table}[b]
\small
\centering
\setlength{\tabcolsep}{7pt}
\caption{Summary Of Dispatch And Curtailment Across All Time Periods For 20\% Wind Penetration And 20 MW BESS}
\label{tab:dispatch values}
\begin{tabularx}{\linewidth}{cccccc}
\hline
\textit{\textbf{Time periods}} & \textit{\textbf{Demand}} & \textit{\textbf{CGD}} & \textit{\textbf{WG}} & \textit{\textbf{BD}} & \multicolumn{1}{c}{\textit{\textbf{WC}}} \\ \hline
T1 & 4598.9 & 4083.7 & 475.1 & 40 & 0 \\ \hline
T2 & 4669.3 & 4159.7 & 476 & 40 & 0 \\ \hline
T3 & 4379.1 & 3885.2 & 453.9 & 40 & 0 \\ \hline
T4 & 4170.6 & 3691.4 & 455.6 & 40 & 0 \\ \hline
T5 & 4143.8 & 3723.3 & 407.1 & 30.4 & 0 \\ \hline
T6 & 4399.4 & 3987.5 & 371.8 & 40 & 0 \\ \hline
T7 & 3603.5 & 3451.6 & 338.7 & -40 & 146.8 \\ \hline
T8 & 3799.1 & 3414.8 & 336.8 & 32.4 & 0 \\ \hline
\end{tabularx}
\footnotesize{\noindent \textbf{Note:} CGD = Conventional generation dispatch, WG = Wind generation, BD = Battery dispatch, WC = Wind curtailment; All values are in MW.}
\end{table}

% \begin{table}[tbhp]
% \small
% \centering
% \setlength{\tabcolsep}{14pt}
% \caption{Wind power curtailments with varying penetration level and batetry size.}
% \label{tab:generator data}
% \begin{tabularx}{\linewidth}{ccccc}
% \hline
% \multirow{2}{*}{\textbf{\begin{tabular}[c]{@{}c@{}}Battery size \\ (MW)\end{tabular}}} & \multicolumn{4}{c}{\textbf{\% Penetration level}} \\
%  & \textit{\textbf{10\%}} & \textit{\textbf{20\%}} & \textit{\textbf{30\%}} & \textit{\textbf{40\%}} \\ \hline
% 0 & 153.3 & 226 & 236 & 282 \\ \hline
% 20 & 73.36 & 146.8 & 157.4 & 202.1 \\ \hline
% 40 & 0 & 66.8 & 76.6 & 122.3 \\ \hline
% 60 & 0 & 0 & 0 & 42.8 \\ \hline
% 80 & 0 & 0 & 0 & 0 \\ \hline
% 100 & 0 & 0 & 0 & 0 \\ \hline
% 120 & 0 & 0 & 0 & 0 \\ \hline
% \end{tabularx}
% \footnotesize{\noindent \textbf{Note:} curtailments values are in MW.}
% \end{table}

% \begin{figure}
%     \centering
%     \includegraphics[width=1\linewidth]{figures/Battery_dispatch.png}
%     \caption{Battery dispatch profile under 20\% wind penetration with 20 MW unit storage capacity.}
%     \label{fig:enter-label}
% \end{figure}

\begin{figure}
    \centering
    \includegraphics[width=1\linewidth]{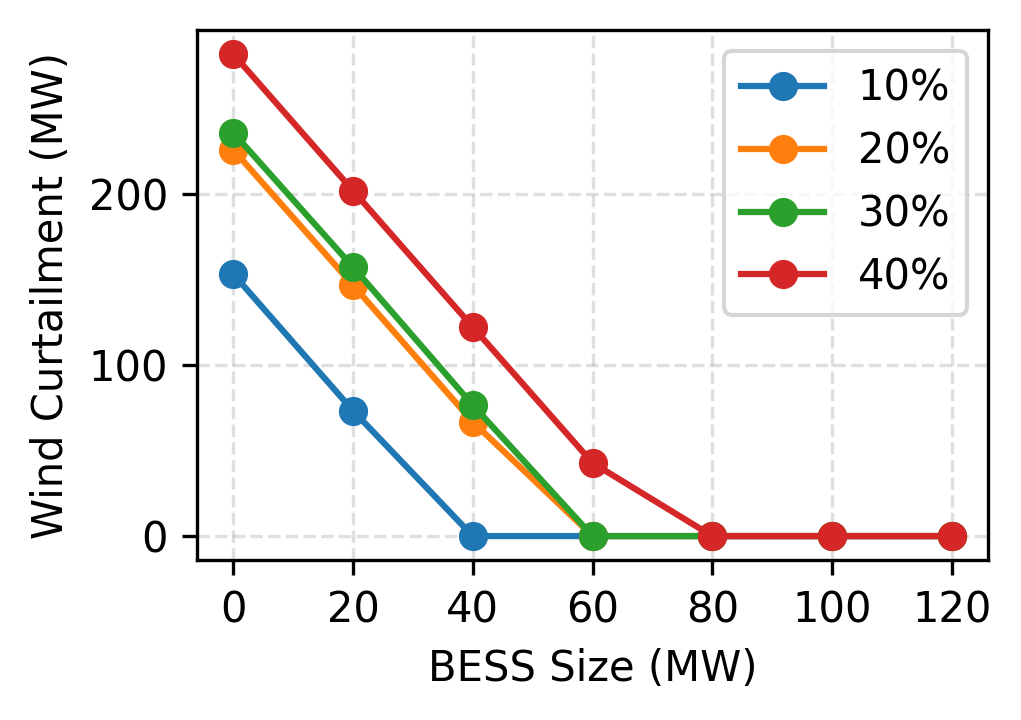}
    \caption{Wind power curtailment as a function of BESS size across different wind penetration levels. }
    \label{fig:wind curtailments with battery size}
\end{figure}

Figure~\ref{fig:wind curtailments with battery size} illustrates wind power curtailments for wind penetration ranging from 10\% to 40\%, as a function of BESS unit size. It is observed that curtailment consistently decreases with larger storage, reaching zero beyond certain thresholds: 40 MW for 10\% penetration, 60 MW for 20\% and 30\%, and 80 MW for 40\%.  Additionally, at smaller BESS sizes, higher wind penetration results in more curtailment. This highlights the reduced operational flexibility in the system as renewable shares increase. These results underscore the importance of appropriately sizing BESS to mitigate curtailments and maintain system flexibility under increasing renewables integration.

\begin{figure}[tbh]
    \centering
    \includegraphics[width=1\linewidth]{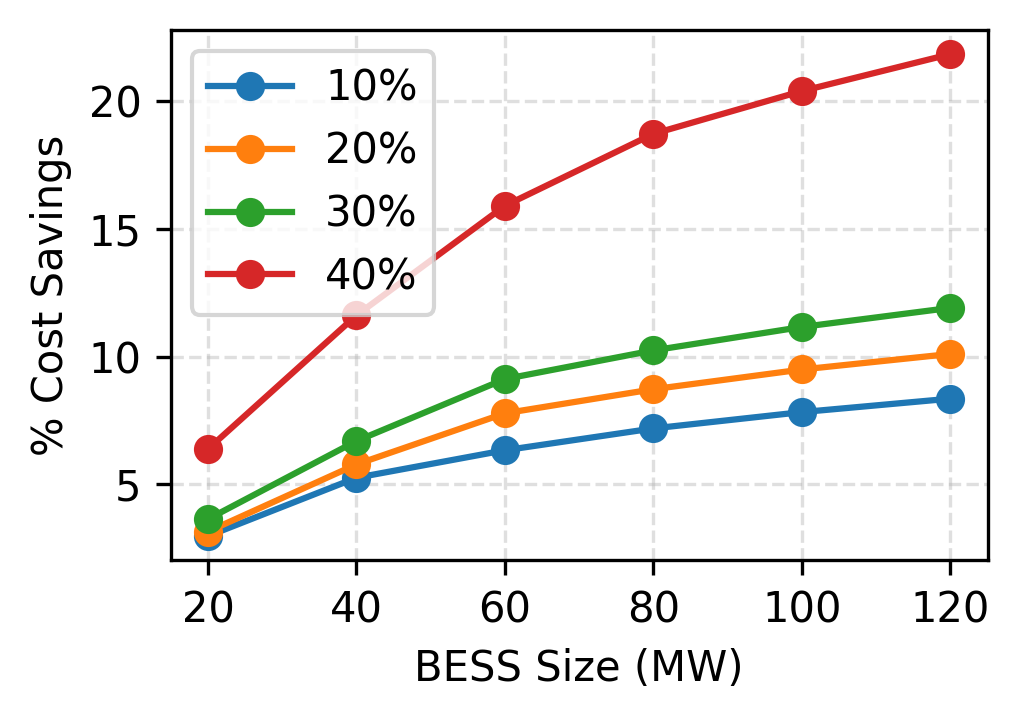}
    \caption{Impact of BESS size on percentage cost savings across different wind penetration levels.}
    \label{fig:impact storage on cost savings}
\end{figure}

Figure~\ref{fig:impact storage on cost savings} illustrates the percentage cost savings achieved with increasing BESS size across different wind penetration levels. The cost saving is calculated relative to a baseline scenario without storage. It is observed that increasing BESS size leads to higher savings for each penetration level, although the marginal benefits diminish at larger capacities. The observation also shows that cost savings increase with higher penetration levels for a given BESS size. It is important to note that a particularly sharp increase in savings is observed when moving from 30\% to 40\% penetration level. For example, at 120 MW BESS capacity, \% savings increase from 11.9\% to 21.8\%, in contrast to the case when penetration increases from 20\% to 30\%. This suggests that beyond a certain level of renewable integration, system flexibility becomes substantially constrained, resulting in higher dispatch costs that can be effectively mitigated by storage. 

\section{Conclusion} \label{sec:Conclusion}

This paper presented a stochastic dynamic economic dispatch with storage (SDED-S) framework to evaluate the impact of battery energy storage system (BESS) in power systems with varying levels of renewables penetration, for which we considered wind power plants. The proposed approach was demonstrated on the IEEE-39 bus system, where selected conventional generators were converted to wind power plants to simulate varying penetration levels. We captured the temporal correlation between wind and load uncertainties and generated scenarios using stratified and importance sampling, where probability distribution data were obtained from NREL. Case studies revealed that strategic BESS deployment significantly enhances system flexibility by reducing the curtailments and dispatch costs. Results showed that curtailments decline with increasing storage size, and becomes zero once BESS size reaches certain thresholds, depending on the penetration levels. Moreover, systems with higher wind penetration level exhibited greater curtailments and higher cost savings from BESS, highlighting the growing need for flexibility as renewable shares increase. Although increasing BESS size improves cost savings, diminishing returns were observed at larger capacities. Future work will focus on scaling the SDED-S framework to larger, more realistic systems with a greater number of scenarios. Decomposition techniques such as Progressive Hedging~\cite{watson2011progressive} and separable projective approximation routine-optimal power flow~\cite{lamichhane2025SPAR} will be explored. Additionally, the framework will be extended to incorporate solar power plants and other uncertainties, such as real-time electricity prices.

\ifCLASSOPTIONcaptionsoff
  \newpage
\fi

\bibliographystyle{IEEEtran}
\end{document}